\documentclass[12pt]{article}

\usepackage{amsmath,amssymb,amsfonts,amsbsy}
\usepackage{cite}
\usepackage{graphicx}
\usepackage{wrapfig}
\usepackage{epsfig}
\usepackage{bbm} % blackboard bold
\usepackage{bm} % blackboard bold
\usepackage{color}                                                       %
\usepackage{dsfont} % double stroke roman fonts                          %
\usepackage{latexsym} % a few additional special symbols                 %
\usepackage{lscape} % allows single pages to be set landscape            %
\usepackage{mathrsfs} % alternative mathematical "script" font           %
\usepackage{morefloats} % solves excess floating figure problems         %
\usepackage{floatflt} % solves excess floating figure problems           %
\usepackage{slashed} % slash notation for Dirac gamma-matrix products    %
\usepackage{psfrag}

\DeclareFontFamily{OT1}{mygreek}{}%
\DeclareFontShape{OT1}{mygreek}{m}{n}{<->omsegr}{}%
\DeclareFontShape{OT1}{mygreek}{b}{n}{<->omsegrb}{}%
\DeclareFontShape{OT1}{mygreek}{m}{it}{<->omsegri}{}%
\DeclareFontShape{OT1}{mygreek}{bx}{n}{<->sub * mygreek/b/n}{}%
\DeclareFontShape{OT1}{mygreek}{m}{sl}{<->sub * mygreek/m/it}{}%
\DeclareSymbolFont{Greekrm}{OT1}{mygreek}{m}{n}
\DeclareSymbolFont{Greekbf}{OT1}{mygreek}{b}{n}
\DeclareSymbolFont{Greekit}{OT1}{mygreek}{m}{it}
\DeclareMathSymbol{\omegab}{\mathalpha}{Greekbf}{119}
\begin{document}
\addcontentsline{toc}{subsection}{{Cross section and single transverse target
> spin asymmetry for backward pion electroproduction}\\
{\it K.~Semenov-Tian-Shansky}}

%%%%%%% please do not touch these! %%%%%%
\setcounter{section}{0}
\setcounter{subsection}{0}
\setcounter{equation}{0}
\setcounter{figure}{0}
\setcounter{footnote}{0}
\setcounter{table}{0}

\begin{center}
\textbf{CROSS SECTION AND SINGLE TRANSVERSE TARGET
 SPIN ASYMMETRY FOR BACKWARD PION ELECTROPRODUCTION}

\vspace{5mm}

B.~Pire$^{\,1  }$, \underline{K.~Semenov-Tian-Shansky}$^{\,1,\,2 \, \dag}$ and
L.~Szymanowski$^{\,3}$

\vspace{5mm}

\begin{small}
  (1) \emph{ CPhT, \'{E}cole Polytechnique, CNRS,  91128, Palaiseau, France} \\
  (2) \emph{IFPA, AGO Dept.,  Universit\'{e} de  Li\`{e}ge, 4000 Li\`{e}ge,  Belgium} \\
  (3) \emph{National Center for Nuclear Research (NCBJ), Warsaw, Poland} \\
  $\dag$ \emph{E-mail: Kirill.Semenov@cpht.polytechnique.fr}
\end{small}
\end{center}

\vspace{0.0mm} % Don't laugh: it does change the spacing!

\begin{abstract}
Nucleon to meson transition distribution amplitudes (TDAs),
non-diagonal  matrix elements of nonlocal three quark operators between a nucleon
and a meson states, arise within the collinear factorized description of hard exclusive
electroproduction of mesons off nucleons in the backward direction.  Below we address the problem of
modelling pion to nucleon TDAs. We suggest a factorized Ansatz for quadruple distributions with input
from the soft pion theorem for
$\pi N$
TDAs. In order to satisfy the polynomiality property in its full form the spectral representation
is complemented with a $D$-term like contribution from the nucleon exchange in the $u$-channel
of the reaction. We present our estimates for the backward pion electroproduction unpolarized cross
section and its transverse target single spin asymmetry within our composite model for
$\pi N$
TDAs.

\end{abstract}

\vspace{7.2mm}

%\section{Introduction}

The possibility to provide a description for hard exclusive electroproduction of
mesons (specifically here pions) off nucleons
\begin{equation}
e(k_1)+ N(p_1) \rightarrow \big( \gamma^*(q) + N(p_1) \big) +e(k_2)
  \rightarrow  e (k_2)+ \pi(p_\pi) + N'(p_2).
\label{reaction}
\end{equation}
in terms of the fundamental degrees of freedom of QCD resides on the collinear factorization theorem
\cite{KSemenov_Collins:1996fb}
valid in the so-called generalized Bjorken limit:
large
$Q^2=-q^2$
and
$s=(p+q)^2$;
fixed
$x_{\rm Bj}= \frac{Q^2}{2 (p \cdot q)}$
and the skewness $\xi$,  defined with respect to the $t$-channel momentum transfer:
%$\Delta \equiv p_2-p_1$:
$ \xi= -\frac{(p_2-p_1) \cdot n}{   (p_1+p_2) \cdot n}$
($n$
is the conventional light cone vector occurring in the Sudakov decomposition of the
relevant momenta) and small $t$-channel momentum transfer squared
$t \equiv (p_2-p_1)^2$.
This factorization theorem allows to present the scattering amplitude as a
convolution of the hard part (coefficient function - CF) with non-perturbative
soft parts (generalized parton distributions - GPDs and distribution amplitudes - DAs)
describing hadronic contents.

\begin{wrapfigure}[10]{R}{40mm}
%% [number of text lines to wrap]{horizontal position: LRC}{figure width}
  \centering %% do not use \begin{center} ... \end{center}
  \vspace*{-8mm} %% the vertical position may need tweaking
  \includegraphics[width=40mm]{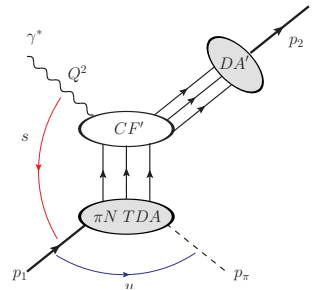}
  \caption{ \footnotesize Factorization for hard production of pions off nucleons in the
     backward kinematics. }
  \label{KSemenov_fig1}
\end{wrapfigure}

According to the conjecture made in
\cite{KSemenov_Frankfurt:1999fp},
a similar collinear factorization theorem for the reaction
(\ref{reaction})
should be valid in the \emph{complementary} kinematical regime:
large $Q^2$
and
$s$;
fixed
$x_{\rm Bj}$
and the skewness variable, which is now defined with respect to
the $u$-channel momentum transfer
$\Delta \equiv p_\pi-p_1$:
$ \xi= -\frac{(\Delta \cdot n)}{  (p_1+p_\pi) \cdot n}$
and small $u$-channel momentum transfer squared
$u \equiv (p_\pi-p_1)^2$
(rather than small
$t$).
Under these assumptions, referred to as the backward kinematics regime,
the amplitude of the reaction
(\ref{reaction})
factorizes as it is presented on  Fig.\ref{KSemenov_fig1}
(see
Ref.~\cite{KSemenov_Lansberg:2007ec}
for the detailed framework).

This requires the introduction of supplementary non-perturbative objects in addition
to GPDs -- nucleon to pion transition distribution amplitudes
($\pi N$ TDAs) defined through the Fourier transform of the
$\pi N$
matrix element of the three-local quark operator on the light cone
\cite{KSemenov_Efremov:1978rn},
\cite{KSemenov_Lepage:1980}:
\begin{equation}
\hat{O}^{\alpha \beta \gamma}_{\rho \tau \chi}( \lambda_1 n,\, \lambda_2 n, \, \lambda_3 n) =
\Psi^\alpha_\rho(\lambda_1 n)
%[\lambda_1 n;\lambda_0 n ]
\Psi^\beta_\tau(\lambda_2 n)
%[\lambda_2 n;\lambda_0 n ]
\Psi^\gamma_\chi (\lambda_3 n).
%[\lambda_3 n;\lambda_0 n ]\,.
\label{oper}
\end{equation}
Here
$\alpha$, $\beta$, $\gamma$
stand for quark flavor indices and
$\rho$, $\tau$, $\chi$
denote the Dirac spinor indices; antisymmetrization in color is implied;  gauge links
are omitted in  the light-like gauge
$A \cdot n=0$.

The physical picture encoded in baryon to meson TDAs is conceptually close to
that contained in baryon GPDs
\cite{KSemenov_Burkardt:2000za}.
By Fourier transforming baryon to meson TDAs to the
impact parameter space
($\Delta_T \to b_T$)
a comprehensible three dimensional physical picture may be obtained.
Baryon to meson TDAs  encode complementary information on the hadron structure in the
transverse plane. In particular, they allow to probe the localization of baryonic charge
in the transverse plane and perform the femto-photography of hadrons
\cite{KSemenov_Ralston:2001xs}
from a new perspective.
There are also hints
\cite{KSemenov_Strikman:2009bd}
that $\pi N$ TDAs
may be used as a tool to perform spatial imaging of the structure of nucleon's  meson cloud.
This  point, which still awaits a detailed exploration,  opens a fascinating window for the
investigation of the various facets of the nucleon's interior.
$\pi N$ TDAs were recently estimated within  the light cone quark model
\cite{KSemenov_Pasquini:2009ki}.

Below we briefly discuss how $\pi N$ TDAs meet the fundamental requirements following  from the
symmetries of QCD, summarizing the main results of
Refs.~\cite{KSemenov_Pire:2010if}, \cite{KSemenov_Pire:2011xv}.
\begin{itemize}
\item  For given flavor contents spin decomposition of the leading twist-$3$
$\pi N$
TDA involve eight invariant functions
$V_{1,2}^{\pi N}$, $A_{1,2}^{\pi N}$, $T_{1,2,3,4}^{\pi N}$
each depending on the  longitudinal momentum fractions
$x_i$ ($\sum_{i=1}^3 x_i=2\xi$),
skewness parameter
$\xi$
and the $u$-channel momentum transfer squared
$\Delta^2 \equiv (p_\pi-p_1)^2$
as well as on the factorization scale $\mu^2$.
\item Not all
$\pi N$
TDAs are independent. Taking the account of the isotopic and permutation symmetries
(see \cite{KSemenov_Pire:2011xv}),
one may check that in order to provide description of all isotopic channels of the reaction
(\ref{reaction})
it suffices to introduce
eight
independent
$\pi N$ TDAs:
four in both  the isospin-$\frac{1}{2}$
and the isospin-$\frac{3}{2}$ channels.

\item The evolution properties of
$\pi N$
TDAs are described by the appropriate generalization
\cite{KSemenov_Pire:2005ax}
of the Efremov-Radyushkin-Brodsky-Lepage/ Dokshitzer-Gribov-Lipatov-Altarelli-Parisi
(ERBL/DGLAP) evolution equations.

\item The support of $\pi N$ TDAs in the longitudinal momentum fractions
$x_i$
is given by the intersection of three stripes
$-1+\xi \le x_i \le 1+\xi$ ($\sum_{i=1}^3 x_i=2\xi$)
\cite{KSemenov_Pire:2010if}.
One can distinguish the ERBL-like domain,
in which all
$x_i$
are positive and two type of DGLAP-like domains, in which
one or two
$x_i$
turn negative.

\item The polynomiality property for the Mellin moments of
$\pi N$
TDAs
in the longitudinal momentum fractions
$x_i$
is the direct consequence of the underlying Lorentz symmetry.
Similarly to the GPD case, the
$(n_1,\,n_2,\,n_3)$-th ($n_1+n_2+n_3 \equiv N$)
Mellin moments of nucleon to meson TDAs  in
$x_1$, $x_2$, $x_3$
are polynomials of power
$N$ or $N+1$
in the skewness variable
$\xi$.

\item Crossing transformation relates
$\pi N$
TDAs to
$\pi N$
generalized distribution
amplitudes (GDAs), defined by the matrix element of the same operator
(\ref{oper})
between the
$\pi N$
state and  the vacuum. The soft pion theorem
\cite{KSemenov_Pobylitsa:2001cz}
for $\pi N$ GDAs
\cite{KSemenov_Braun}
also constrains
$\pi N$
TDAs near the soft pion threshold
$\xi=1$, $\Delta^2=M^2$, where $M$ is the nucleon mass.

\end{itemize}

%\section{Spectral representation for $\pi N$ TDAs}

The most direct way to ensure both the polynomiality and the support properties for
$\pi N$
TDAs is to employ the spectral representation
in terms of quadruple distributions
\cite{KSemenov_Pire:2010if}.
Our strategy of modeling
$\pi N$
TDAs
\cite{KSemenov_Paper3}
is completely analogous to that employed for modeling nucleon
GPDs with the help of Radyushkin's double distribution Ansatz
\cite{KSemenov_RDDA4}.
The main difficulty is that, contrary to GPDs, baryon to meson TDAs lack a comprehensible
forward limit ($\xi=0$).
In order to propose a model for quadruple distributions
it is  illuminating to consider the alternative limit
$\xi=1$,
in which
$\pi N$ TDAs are constrained by the chiral dynamics through the  soft pion theorem
\cite{KSemenov_Pobylitsa:2001cz}
for $\pi N$ GDAs.
In this limit $\pi N$ TDAs are
expressed through the nucleon DAs
$\{V^p,\,A^p,\,T^p\}$
\cite{KSemenov_Chernyak_Nucleon_wave}.
For example,
$\pi N$
TDAs
$V^{\pi^0 p}_1$, $A^{\pi^0 p}_1$, $T^{\pi^0 p}_1$
reduce to the following combination of the nucleon DAs
\cite{KSemenov_Pire:2011xv}:
\begin{eqnarray}
&&
\big\{ V^{\pi^0 p}_1, \, A^{\pi^0 p}_1 \big\}(x_1,x_2,x_3,\xi=1)= -\frac{1}{4} \times \frac{1}{2} \big\{ V^{p}, \, A^{p}
\big\}
\left(\frac{x_1}{2},\frac{x_2}{2},\frac{x_3}{2}\right);
\nonumber \\ &&
T^{\pi^0 p}_1(x_1,x_2,x_3,\xi=1)= \frac{1}{4} \times \frac{3}{2} T^{p}\left(\frac{x_1}{2},\frac{x_2}{2},\frac{x_3}{2}\right),
\label{Soft_pion_th_pi0p}
\end{eqnarray}

and
 $\big\{ V^{\pi^0 p}_2, \, A^{\pi^0 p}_2, \, T^{\pi^0 p}_2 \big\}(x_1,x_2,x_3,\xi=1)=-\frac{1}{2}
\big\{ V^{\pi^0 p}_1, \, A^{\pi^0 p}_1, \, T^{\pi^0 p}_1 \big\}(x_1,x_2,x_3,\xi=1)$.

With appropriate change of spectral parameters
the spectral representation for
$\pi N$ TDAs of Ref.~\cite{KSemenov_Pire:2010if}
can be  rewritten as:
\begin{eqnarray}
&&
H (w_i,\,v_i,\,\xi)=
\int_{-1}^1 d \kappa_i \int_{- \frac{1-\kappa_i}{2}}^{ \frac{1-\kappa_i}{2}} d\theta_i
\int_{-1}^1 d \mu_i \int_{- \frac{1-\mu_i}{2}}^{ \frac{1-\mu_i}{2}} d\lambda_i
\, \delta \big(w_i- \frac{\kappa_i-\mu_i}{2} (1-\xi) - \kappa_i \xi\big)  \nonumber \\ &&
\times
\delta\big(v_i- \frac{\theta_i-\lambda_i}{2} (1-\xi) - \theta_i \xi \big) \,
F(\kappa_i, \, \theta_i, \mu_i,\, \lambda_i ).
\label{Spectral_for_TDAs_redy_to_be_factorized}
\end{eqnarray}
The index $i=1,2,3$ here refers to one of three possible choices of independent variables
(quark-diquark coordinates):
$w_i= x_i-\xi$, $v_i= \frac{1}{2} \sum_{k,l=1}^3 \varepsilon_{ikl} x_k$.
We suggest to use the following factorized Ansatz for the quadruple distribution
$F$
in
(\ref{Spectral_for_TDAs_redy_to_be_factorized}):
\begin{eqnarray}
F(\kappa_i, \, \theta_i,\, \mu_i,\, \lambda_i )= 4 V(\kappa_i, \, \theta_i) \, h(\mu_i,\, \lambda_i),
\label{Factorized_ansatz_xi=1}
\end{eqnarray}
where
$V(\kappa_i, \, \theta_i)$
is the combination of nucleon DAs
%$V(y_1,y_2,y_3)$ ($\sum_{i=1}^3 y_i=1$)
to which
$\pi N$
TDA in question
reduces in the limit
$\xi=1$
({\it c.f.} Eq.~(\ref{Soft_pion_th_pi0p})),
rewritten in terms of independent variables:
$\kappa_i= 2y_i-1$;
$\theta_i=  \sum_{k,l=1}^3 \varepsilon_{ikl} y_k$.

The profile function
$h(\mu_i,\, \lambda_i)$
is  normalized as
$
\int_{-1}^1 d \mu_i \int_{- \frac{1-\mu_i}{2}}^{ \frac{1-\mu_i}{2}} d\lambda_i \, h(\mu_i,\, \lambda_i) =1\,.
$
The support of the profile function
$h$ is also that of a baryon DA.
The simplest assumption for the profile is to take it to  be determined by the
asymptotic form of baryon DA
($120 y_1 y_2 y_3$
with
$\sum_{i=1}^3 y_i=1$)
rewritten in terms of variables
$\mu_i= 2y_i-1$, $\lambda_i=   \sum_{k,l=1}^3 \varepsilon_{ikl} y_k$:
\begin{equation}
h(\mu_i,\, \lambda_i)=
\frac{15}{16} \, (1+\mu_i) ((1-\mu_i)^2-4 \lambda_i^2).
\label{Profile}
\end{equation}

Similarly to the GPD case
\cite{KSemenov_Polyakov:1999gs},
in order to satisfy the polynomiality condition in its complete form the spectral
representation for
$\pi N$
TDAs
$\{V_{1,2}, \,A_{1,2},\, T_{1,2} \}^{\pi N}$
should be complemented by a $D$-term like contribution.
The simplest possible model for such a $D$-term
is the contribution of the $u$-channel nucleon exchange
into $\pi N$ TDAs computed in
\cite{KSemenov_Pire:2011xv}.
In this way we come to a two component model for
$\pi N$
TDAs involving the spectral representation part, based on the factorized Ansatz
(\ref{Factorized_ansatz_xi=1}) with the profile
(\ref{Profile})
and with input from
the soft pion theorem,
and the $D$-term, originating from the nucleon exchange in the $u$-channel.
It provides a model for $\pi N$ TDAs in the complete domain of their definition.

Within the factorized approach the leading order (both in
$\alpha_s$
and
$1/Q$)
amplitude of backward hard pion production
$\mathcal{M}^\lambda_{s_1s_2}$
reads \cite{KSemenov_Lansberg:2007ec}:
\begin{eqnarray}
{\mathcal M}^\lambda_{s_1s_2}=\mathcal{C} \frac{1}{Q^4} \Big[
{\cal S}^\lambda_{s_1s_2} \mathcal{I}(\xi,\Delta^2)+
{\cal S}'^\lambda_{s_1s_2} \mathcal{I}'(\xi,\Delta^2) \Big].
\label{helicity_ampl_rewr}
\end{eqnarray}
The spin structures
${\cal S}$
and
${\cal S}'$
are defined as
$
{\cal S}^\lambda_{s_1s_2} \equiv \bar U(p_2,s_2)
\hat{\mathcal{E}}(\lambda)
 \gamma^5 U(p_1,s_1)\,$;
 %%%%%%%%%%%%%%%%%%%%%%%%%%%%%%%%
   ${\cal S'}^\lambda_{s_1s_2} \equiv \frac{1}{M}\bar U(p_2,s_2)
\hat{ \mathcal{E} }(\lambda) \hat{\Delta}_T
 \gamma^5 U(p_1,s_1),
$
where
$\mathcal{E}$
denotes the polarization vector of the virtual photon and
$U$
is the usual nucleon Dirac spinor.
$\mathcal{C}$ is the normalization constant
 $\mathcal{C} \equiv
-i
\frac{(4 \pi \alpha_s)^2 \sqrt{4 \pi \alpha_{em}} f_{N}^2}{ 54 f_{\pi} }$,
where
$\alpha_{em} (\alpha_s)$
stands for the electromagnetic (strong) coupling,
$f_\pi=93$MeV
is the pion decay constant and
$f_N$
is the normalization constant of the nucleon DA \cite{KSemenov_Chernyak_Nucleon_wave}.

The coefficients
$\mathcal{I}$, $\mathcal{I}'$
result from the calculation
of $21$ diagrams contributing to the hard scattering amplitude
\cite{KSemenov_Lansberg:2007ec}:
\begin{equation*}
%&&
\big\{ \mathcal{I}, \, \mathcal{I}' \big\}(\xi,\Delta^2) \equiv \! \! {\int \! d^3 x
\delta(\sum_i x_i-2\xi)
\int \! d^3y \delta(\sum_i y_i-1)
\Bigg(2\sum_{\alpha=1}^{7}  \big\{T_{\alpha}, \, T_{\alpha}'\big\}+
\sum\limits_{\alpha=8}^{14} \big\{T_{\alpha}, \, T_{\alpha}'\big\}\Bigg)},
\label{Def_IIprime}
\end{equation*}
where the convolution integrals in
 $x_i$
 and
 $y_i$
stand over the supports of
$\pi N$
TDAs and nucleon DAs respectively.
The explicit expressions for the coefficients
$T_\alpha$ and $T'_\alpha$
for
$\gamma^\star p \rightarrow \pi^0 p$ channel
are presented in the Table~I of Ref.~\cite{KSemenov_Lansberg:2007ec}.
The result for
$\gamma^\star p \rightarrow \pi^+ n$ channel can be read off the same
Table
with the obvious changes:
$Q^u  \rightleftarrows Q^d$;
$\{V_{1,2},A_{1,2},T_{1,2,3,4}\}^{p \pi^0} \rightarrow \{V_{1,2},A_{1,2},T_{1,2,3,4}\}^{p \pi^+}$;
$\{V,\,A,\,T\}^p \rightarrow \{V,\,A,\,T\}^n$.
In
\cite{KSemenov_Paper3}
we develop a reliable method for the calculation of the corresponding
convolution integrals.

\begin{wrapfigure}[10]{R}{60mm}
%% [number of text lines to wrap]{horizontal position: LRC}{figure width}
  \centering %% do not use \begin{center} ... \end{center}
  \vspace*{-8mm} %% the vertical position may need tweaking
  \includegraphics[width=60mm]{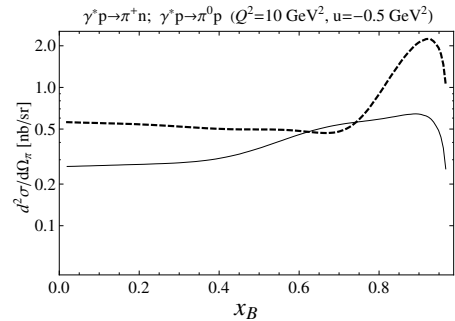}
  \caption{\footnotesize
Unpolarized cross section
  for backward $\pi^+$  (solid) and $\pi^0$ (dashed) production off proton. }
  \label{KSemenov_fig3}
\end{wrapfigure}

Within the suggested factorization mechanism for backward pion electroproduction only the
transverse cross section
$\frac{d^2 \sigma_T}{d \Omega_\pi}$
receives a contribution at the leading twist level.
We establish the following formula for the unpolarized transverse cross section
through the coefficients
$\mathcal{I}$, $\mathcal{I}'$
introduced in
(\ref{helicity_ampl_rewr}):
\begin{eqnarray}
\frac{d^2 \sigma_T}{d \Omega_\pi}= |\mathcal{C}|^2 \frac{1}{Q^6}
\frac{\Lambda(s,m^2,M^2)}{128 \pi^2 s (s-M^2)} \frac{1+\xi}{\xi}
\big(
|\mathcal{I}|^2
-  \frac{\Delta_T^2}{M^2} |\mathcal{I}'|^2
\big),
\label{Work_fla_CS}
\end{eqnarray}
where
$
\Lambda(x,y,z)= \sqrt{x^2+y^2+z^2-2xy-2xz-2yz}
$
is the usual  Mandelstam function.
Within our two component model for $\pi N$ TDAs
$\mathcal{I}$ receives contributions both
from the spectral representation component and nucleon pole exchange contribution.
$\mathcal{I}'$ is determined solely by the nucleon pole contribution.
On Fig.~\ref{KSemenov_fig3}
we present our estimates for the unpolarized cross section
$\frac{d^2 \sigma_T}{d \Omega_\pi}$
of backward production of
$\pi^+$
and
$\pi^0$
off protons for
$Q^2=10 {\rm GeV}^2$
and $u=-0.5 {\rm GeV}^2$
in ${\rm nb}/{\rm sr}$.
CZ solution \cite{KSemenov_Chernyak_Nucleon_wave}
for the nucleon DAs is used as phenomenological  input  for our model. The magnitude of
the  cross sections is large enough for a detailed investigation to be carried
at high luminosity experiments such as J-lab@12GeV and EIC.
The scaling law for the unpolarized \\ cross section (\ref{Work_fla_CS}) is
$1/Q^8$.

%%%%%%%% Fig. 3 %%%%%%%%
%%%%%%%%%%%%%%%%%%%%%%%%

Asymmetries, being ratios of the cross sections, are less sensitive to perturbative
corrections. Therefore, they are usually considered as  more reliable observables to
test the factorized description of hard reactions. For the backward pion
electroproduction the evident candidate is the single transverse target
spin asymmetry (STSA)
\cite{KSemenov_Lansberg:2010mf}
defined as:
\begin{eqnarray}
&&
\mathcal{A}= \frac{1}{|\vec{s}_1|}
\frac{\left(
\int_0^\pi d \tilde{\varphi} |\mathcal{M}_{T}^{s_1}|^2 - \int_\pi^{2\pi} d \tilde{\varphi} |\mathcal{M}_{T}^{s_1}|^2
\right)}{ \left(
\int_0^{2\pi} d \tilde{\varphi} |\mathcal{M}_{T}^{s_1}|^2
\right) }
 \nonumber \\ &&
 = -\frac{4}{\pi} \frac{\frac{|\Delta_T|}{M}  \, {\rm Im} (\mathcal{I}'(\mathcal{I})^*)}{|\mathcal{I}|^2
- \frac{\Delta_T^2}{M^2} |\mathcal{I}'|^2}.
\label{Def_asymmetry}
\end{eqnarray}
Here $\tilde \varphi \equiv \varphi -\varphi_s$, where
$\varphi$ is the angle between leptonic and hadronic planes and $\varphi_s$ is the angle between
the leptonic plane and the transverse spin of the target.
Our two component model for
$\pi N$
TDAs provides a non-vanishing  numerator in the last equality of
(\ref{Def_asymmetry})
 thanks to the  interfering contributions of the spectral part into
${\rm Im} \mathcal{I}(\xi)$
and of the nucleon pole part into
${\rm Re} \mathcal{I}'(\xi)$.

\begin{wrapfigure}[10]{R}{60mm}
%% [number of text lines to wrap]{horizontal position: LRC}{figure width}
  \centering %% do not use \begin{center} ... \end{center}
  \vspace*{-8mm} %% the vertical position may need tweaking
  \includegraphics[width=60mm]{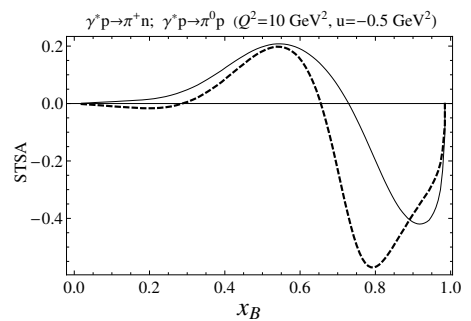}
  \caption{\footnotesize
 Single transverse spin asymmetry
  for backward $\pi^+$ (solid) and  $\pi^0$  (dashed) production off proton. }
  \label{KSemenov_fig4}
\end{wrapfigure}

On Fig.~\ref{KSemenov_fig4} we show the result of our calculation of the
STSA for backward
$\pi^+$
and
$\pi^0$
electroproduction off protons for
$Q^2=10 \, {\rm GeV}^2$ and $u=-0.5 \, {\rm GeV}^2$.
STSA turns out to be sizable in the valence region and its measurement should   be
considered as a crucial test of the applicability of our collinear factorized \\ scheme  for
backward pion electroproduction.

%%%%%%%% Fig. 3 %%%%%%%%
%%%%%%%%%%%%%%%%%%%%%%%%
Our estimates of backward pion electroproduction cross section and single transverse spin asymmetry make
us hope for bright experimental prospects for measuring baryon to meson TDAs with high luminosity electron
beams such   as J-lab@ 12 GeV and EIC.
Experimental data from J-lab@ 6 GeV on backward
$\pi^+$, $\pi^0$, $\eta$
and
$\omega$
meson production are currently being analyzed
\cite{KSemenov_J-lab}.
We eagerly await for the experimental evidences for
validity of the  factorized picture  of backward electroproduction
reactions suggested in our approach.

\vspace{.1cm}

This work is supported in part by the Polish NCN grant DEC-2011/01/B/ST2/03915
and by the French-Polish Collaboration Agreement Polonium.


\begin{thebibliography}{99} %% for less than 10 references use just {9}

\bibitem{KSemenov_Collins:1996fb}
  J.~C.~Collins, L.~Frankfurt and M.~Strikman,
  %``Factorization for hard exclusive electroproduction of mesons in QCD,''
  Phys.\ Rev.\  D {\bf 56} (1997) 2982.
 % [arXiv:hep-ph/9611433].
  %%CITATION = PHRVA,D56,2982;%%

 \bibitem{KSemenov_Frankfurt:1999fp}
  L.~L.~Frankfurt, P.~V.~Pobylitsa, M.~V.~Polyakov and M.~Strikman,
  %``Hard exclusive pseudoscalar meson electroproduction and spin structure  of
  %a nucleon,''
  Phys.\ Rev.\  D {\bf 60} (1999) 014010;
%  [arXiv:hep-ph/9901429];
  L.~Frankfurt, M.~V.~Polyakov, M.~Strikman, D.~Zhalov and M.~Zhalov,
  %``Novel hard semiexclusive processes and color singlet clusters in hadrons,''
  arXiv:hep-ph/0211263.
  %%CITATION = HEP-PH/0211263;%%

\bibitem{KSemenov_Efremov:1978rn}
  A.~V.~Efremov and A.~V.~Radyushkin,
  %``Asymptotical Behavior Of Pion Electromagnetic Form-Factor In QCD,''
  Theor.\ Math.\ Phys.\  {\bf 42} (1980) 97
  [Teor.\ Mat.\ Fiz.\  {\bf 42} (1980) 147].
  %%CITATION = TMFZA,42,147;%%

%\cite{Lepage:1979zb}
\bibitem{KSemenov_Lepage:1980}
  G.~P.~Lepage and S.~J.~Brodsky, Phys. Rev. D {\bf 22} (1980) 2157.

\bibitem{KSemenov_Lansberg:2007ec}
  J.~P.~Lansberg, B.~Pire and L.~Szymanowski,
  %``Hard exclusive electroproduction of a pion in the backward region,''
  Phys.\ Rev.\  D {\bf 75} (2007) 074004
  [Erratum-ibid.\  D {\bf 77} (2008) 019902].

\bibitem{KSemenov_Burkardt:2000za}
  M.~Burkardt,
  %``Impact parameter dependent parton distributions and off-forward parton
  %distributions for zeta --> 0,''
  Phys.\ Rev.\  D {\bf 62} (2000) 071503.
  %[Erratum-ibid.\  D {\bf 66}, 119903 (2002)]
  %[arXiv:hep-ph/0005108].

\bibitem{KSemenov_Ralston:2001xs}
  J.~P.~Ralston and B.~Pire,
  %``Femto-photography of protons to nuclei with deeply virtual Compton
  %scattering,''
  Phys.\ Rev.\  D {\bf 66} (2002) 111501.
 % [arXiv:hep-ph/0110075].

 %\cite{Strikman:2009bd}
\bibitem{KSemenov_Strikman:2009bd}
  M.~Strikman and C.~Weiss,
  %``Chiral dynamics and partonic structure at large transverse distances,''
  Phys.\ Rev.\  D {\bf 80} (2009) 114029.
 % [arXiv:0906.3267 [hep-ph]].
  %%CITATION = PHRVA,D80,114029;%%

\bibitem{KSemenov_Pasquini:2009ki}
 M.~Pincetti, B.~Pasquini and S.~Boffi,
  %``Nucleon to Pion Transition Distribution Amplitudes in a Light-Cone Quark Model,''
  arXiv:0807.4861 [hep-ph];
  B.~Pasquini, M.~Pincetti and S.~Boffi,
  %``Parton content of the nucleon from distribution amplitudes and transition
  %distribution amplitudes,''
  Phys.\ Rev.\  D {\bf 80} (2009) 014017.
 % [arXiv:0905.4018 [hep-ph]].
  %%CITATION = PHRVA,D80,014017;%%

\bibitem{KSemenov_Pire:2010if}
   B.~Pire, K.~Semenov-Tian-Shansky and L.~Szymanowski,
  %``A Spectral representation for baryon to meson transition distribution
  %amplitudes,''
  Phys.\ Rev.\  D {\bf 82} (2010) 094030.
 % [arXiv:1008.0721 [hep-ph]].

\bibitem{KSemenov_Pire:2011xv}
B.~Pire, K.~Semenov-Tian-Shansky, L.~Szymanowski,
  %``\pi N transition distribution amplitudes: their symmetries and constraints from chiral dynamics,''
  Phys.\ Rev.\ D {\bf 84} (2011) 074014.
  %[arXiv:1106.1851 [hep-ph]].

\bibitem{KSemenov_Pire:2005ax}
  B.~Pire and L.~Szymanowski,
  %``QCD analysis of anti-p N --> gamma* pi in the scaling limit,''
  Phys.\ Lett.\  B {\bf 622} (2005) 83.
  %[arXiv:hep-ph/0504255].
  %%CITATION = PHLTA,B622,83;%%

%%%%%%%%%%%%%%%%%%%%%%%%%%%%%%%%%%%%%%%%%%%%%%%%%%%%%%%%%%%%%%%%%%%
%                                                                 %
%                  Soft pion theorem                              %
%%%%%%%%%%%%%%%%%%%%%%%%%%%%%%%%%%%%%%%%%%%%%%%%%%%%%%%%%%%%%%%%%%%

%\cite{Pobylitsa:2001cz}
\bibitem{KSemenov_Pobylitsa:2001cz}
  P.~V.~Pobylitsa, M.~V.~Polyakov and M.~Strikman,
  %``Soft pion theorems for hard near threshold pion production,''
  Phys.\ Rev.\ Lett.\  {\bf 87} (2001) 022001.
% [arXiv:hep-ph/0101279].
  %%CITATION = PRLTA,87,022001;%%


 %\cite{hep-ph/0611386}
\bibitem{KSemenov_Braun}
  V.~M.~Braun, D.~Y.~Ivanov, A.~Lenz and A.~Peters,
  %``Deep inelastic pion electroproduction at threshold,''
  Phys.\ Rev.\ D\ {\bf 75} (2007) 014021.
 % [hep-ph/0611386].
  %%CITATION = PHRVA,D75,014021;%%


\bibitem{KSemenov_Paper3}
J.~P.~Lansberg, B.~Pire, K.~Semenov-Tian-Shansky and L.~Szymanowski,
  %``A consistent model for \pi N transition distribution amplitudes and backward pion electroproduction,''
  arXiv:1112.3570 [hep-ph].


\bibitem{KSemenov_RDDA4}
  I.~V.~Musatov and A.~V.~Radyushkin,
  %``Evolution and models for skewed parton distributions,''
  Phys.\ Rev.\  D {\bf 61} (2000) 074027.
  %[arXiv:hep-ph/9905376].

\bibitem{KSemenov_Chernyak_Nucleon_wave}
V.L. Chernyak and I.R. Zhitnitsky, Nucl. Phys. {\bf B 246} (1984) 52.

%%%%%%%%%%%%%%%%%%%%%%%%%%%%%%%%%%%%%%%%%%%%%%%%%%%%%%%%%%%%%%%%%%%
%               D-term                                            %
%%%%%%%%%%%%%%%%%%%%%%%%%%%%%%%%%%%%%%%%%%%%%%%%%%%%%%%%%%%%%%%%%%%
%\cite{Polyakov:1999gs}
\bibitem{KSemenov_Polyakov:1999gs}
  M.~V.~Polyakov and C.~Weiss,
  %``Skewed and double distributions in pion and nucleon,''
  Phys.\ Rev.\  D {\bf 60} (1999) 114017.
 % [arXiv:hep-ph/9902451].
  %%CITATION = PHRVA,D60,114017;%%




%\cite{Lansberg:2010mf}
\bibitem{KSemenov_Lansberg:2010mf}
  J.~P.~Lansberg, B.~Pire, L.~Szymanowski,
  %``Spin Observables in Transition-Distribution-Amplitude Studies,''
  J.\ Phys.\ Conf.\ Ser.\  {\bf 295} (2011) 012090.
  %[arXiv:1011.6635 [hep-ph]].





\bibitem{KSemenov_J-lab}
K.~Park, V.~Kubarovsky, P.~Stoler,  {\it private communication}.





\end{thebibliography}
\end{document}